\documentclass[twocolumn,showpacs,preprintnumbers,amsmath,amssymb]{revtex4}

\usepackage{amsfonts}
\usepackage{graphics}
\usepackage{psfrag}
\usepackage{hyperref}
\usepackage{graphicx}
\usepackage{epsfig}
\usepackage[latin1]{inputenc}
\usepackage{color}
\usepackage{rotating}
\usepackage{multirow}

\usepackage{graphicx}
\usepackage{dcolumn}
\usepackage{bm}

\begin{document}
\title{Building reputation systems for better ranking}

\author{Luo-Luo Jiang$^{1,2}$}
\author{Mat\'u\v{s} Medo$^{1}$}
\author{Joseph R. Wakeling$^{1}$}
\author{Yi-Cheng Zhang$^{1}$}
\author{Tao Zhou$^{1,2,3}$}
\email{zhutou@ustc.edu}

\affiliation{$^1$Department of Physics, University of Fribourg,
Chemin du Muse, CH-1700 Fribourg, Switzerland \\ $^2$Department of
Modern Physics, University of Science and Technology of China, Hefei
230026, PR China \\ $^3$Web Sciences Center, University of
Electronic Science and Technology of China, 610054 Chengdu, PR
China}

\date{\today}

\begin{abstract}
How to rank web pages, scientists and online resources has recently
attracted increasing attention from both physicists and computer
scientists. In this paper, we study the ranking problem of rating
systems where users vote objects by discrete ratings. We propose an
algorithm that can simultaneously evaluate the user reputation and
object quality in an iterative refinement way. According to both the
artificially generated data and the real data from MovieLens and
Amazon, our algorithm can considerably enhance the ranking accuracy.
This work highlights the significance of reputation systems in the
Internet era and points out a way to evaluate and compare the
performances of different reputation systems.
\end{abstract}


\maketitle

\section{Introduction}
Ranking may not be the best way to describe a system, but definitely
provides valuable and impressive information, especially for the
people who do not comprehensively understand the internal
interactions and organization of this system. Nowadays, ranking
techniques are becoming increasingly important in many online
services, and we are always curious for rankings of web pages,
books, scientists, movies, movie stars, and so on. For a simple
undirected graph, the centralities are usually used to rank the
importance of nodes \cite{Freeman2004}, while for directed graph,
PageRank is the most widely applied algorithm who mimics the random
walk process with restart \cite{Brin1998}. Considering a possibly
underlying mixing role of each node, the HITS algorithm
\cite{Kleinberg1999} may provide better ranking. Recently, some
scientists proposed a number of iterative refinement algorithms to
rank the scientists and scientific publications based on the
citation and co-authorship data
\cite{Walker2007,Ding2009,Radicchi2009}.

In this paper, we consider the ranking problem in a different kind
of systems called the rating systems, where each user vote some
objects with ratings (usually discrete ratings from 1 to 5, like in
Netflix.com and Amazon.com). A straightforward method is to rank
objects according to their average ratings. However, a drawback is
that some users are not serious to their votes at all, therefore the
evaluation by simply averaging all ratings may be less accurate. A
promising way to overcome this problem is to estimate the reputation
or trust of each user and to assign more weight to the user with
higher reputation. In fact, to build reputation systems or
reputation societies is a vital task in the Internet era
\cite{Masum2004}, which could find its applications in personalized
recommendations \cite{Massa2004,Ziegler2004}, management of peer to
peer systems \cite{Gupta2003,Wei2006,Walsh2006}, online sales in
e-commerce systems \cite{Resnick2002,Josang2007}, design of mobile
ad-hoc networks \cite{Buchegger2003}, and so on. However, to
estimate the reputation of a user is not a trivial task. Yu \emph{et
al.} \cite{Yu2006,Laureti2006} proposed an iterative refinement
algorithm, where the quality of an object is quantified by its
weighted average rating and a user whose ratings are closer to the
weighted average ratings is considered to be of higher reputation. A
user having higher reputation will be assigned more weight. At each
time step, every user's reputation and every object's weighted
average rating are recalculated, until the system converges to
steady distributions of reputations and weighted average ratings. To
achieve better estimation of user reputation and object quality, the
basic iterative refinement model can be further extended by
accounting for the truncation of the rating and by assuming a prior
distribution on the parameters according to a Bayesian model
\cite{Fouss2010}. Similar problems based on partial information
\cite{Laureti2005} and changing data \cite{Kerchove2007} have also
been considered.

Most of the previous works used artificially generated data to
evaluate the algorithmic performance. In this paper, beyond the
artificial data, we use real data to test a modified iterative
algorithm. The winners of the \emph{Best Picture of Oscar Awards}
among the movies in MovieLens data and the winners of the
\emph{National Book Awards} among the books in Amazon.com are
treated as benchmark objects. Experimental analysis shows that our
modified algorithm gives considerably higher ranks of the benchmark
objects than the average ratings.

\begin{figure}
\begin{center}
\includegraphics*[scale=1.2]{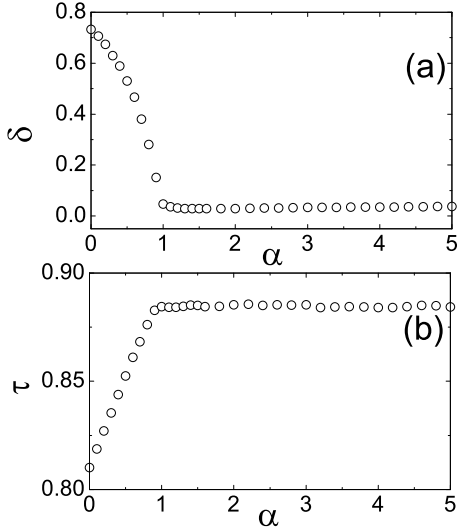}
\end{center}
\caption{$\delta$ and $\tau$ as functions of $\alpha$, where $Q$ and
$\zeta$ obey the uniform distribution. The rating density is fixed
as $\rho=0.05$. All data points are obtained by averaging 100
independent realizations.}
\end{figure}

\section{Method}
A rating system consists of $N$ users and $M$ objects, where each
user rates some objects. Denoting by $\rho$ ($0\leq \rho \leq 1$)
the density of ratings (each user has voted $\rho M$ objects on
average), $x_{ik}$ the rating of object $k$ by user $i$, and $Q_{k}$
the intrinsic quality of object $k$ which is usually not observable.
If $Q_k$ is known, the mean square deviation of user $i$'s votes
from the objects' intrinsic qualities is:
\begin{equation}
\sigma_{i}=\frac{1}{M_{i}}\sum_{k}(x_{ik}-Q_{k})^{2},
\end{equation}
where $k$ runs over all the $M_i$ objects voted by user $i$. We
assume that the user with higher reputation has averagely smaller
$\sigma$, namely higher reputation corresponds to better judgement
of the intrinsic qualities of objects. However, the intrinsic
qualities can not be observed directly, and thus we can only
estimate them based on the users' ratings. Instead of simply
averaging over all ratings, in our reputation system, we assign
higher opinion weight to the user with higher reputation. Denoting
by $\xi_i$ the mean square deviation and thus $\xi_i^{-1}$ the
reputation of user $i$, we assign a weight $\xi_i^{-\alpha}$ to user
$i$ with $\alpha\geq 0$ a free parameter, and thus the estimated
quality of object $k$, measured by the weighted average rating, is
\begin{equation}
q_{k}=\frac{1}{N_k}\sum_ix_{ik}\cdot\frac{\xi_{i}^{-\alpha}}{\sum_j\xi_{j}^{-\alpha}},
\end{equation}
where $N_k$ is the number of users having voted object $k$ and $i$,
$j$ run over all these $N_k$ users. At the same time, the mean
square deviation of user $i$'s ratings can be estimated as
\begin{equation}
\xi_{i}=\frac{1}{M_{i}}\sum_k(x_{ik}-q_k)^{2},
\end{equation}
where $k$ runs over all the $M_i$ objects voted by user $i$. When
$\xi_i<10^{-5}$, we set $\xi_i=10^{-5}$ to avoid divergence.

Equations (2) and (3) describe an iterative refinement method to
estimate the user reputation and object quality. We set the initial
condition as $\forall_i\xi_i=1$, and at each time step we first
estimate $q_k$ by Eq. (2) and then update $\xi_i$ by Eq. (3). The
maximal difference for $\vec{q}$ and $\vec{\xi}$ at the $n$th time
step is defined as:
\begin{equation}
\Delta q(n)=\max_k|q_k(n)-q_k(n-1)|,
\end{equation}
\begin{equation}
\Delta \xi(n)=\max_i|\xi_i(n)-\xi_i(n-1)|.
\end{equation}
The iterative process stops when both $\Delta q$ and $\Delta \xi$
are smaller than the threshold $\Delta_c=10^{-5}$, and the resulted
$\vec{q}$ and $\vec{\xi}$ are used to rank the object quality and
user reputation, respectively.

\begin{figure}
\begin{center}
\includegraphics*[scale=1.2]{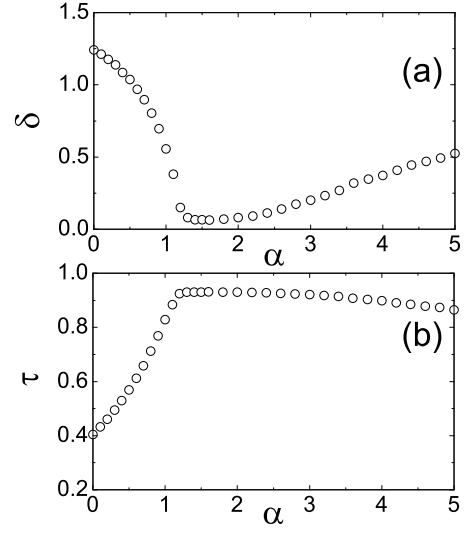}
\end{center}
\caption{$\delta$ and $\tau$ as functions of $\alpha$, where $Q$
obeys the power-law distribution $p(Q)\sim Q^{-1.5}$ and $\zeta$
obeys the uniform distribution. The rating density is fixed as
$\rho=0.05$. All data points are obtained by averaging 100
independent realizations.}
\end{figure}

\section{Results of Artificial Data}
In this section, we test our algorithm by artificial system where
the numbers of users and objects are fixed as $N=2000$ and $M=1000$.
We first generate the intrinsic qualities of objects $\vec{Q}$ and
the noise levels of users' judgements $\vec{\zeta}$ according to
some given distributions (see later). Here the known (exact)
qualities and mean square derivations are denoted by $\vec{Q}$ and
$\vec{\sigma}$ (later we will see that in the statistical level
$\sigma_i\sim \zeta_i^2$), while the estimated values are $\vec{q}$
and $\vec{\xi}$. Then for each user-object ($i-k$) pair, with
probability $\rho$, we generate the artificial rating $x_{ik}$ as
\begin{equation}
x_{ik}=Q_k+\psi \zeta_i,
\end{equation}
where $\psi\in [-1,1]$ is a random variable. The lower and upper
boundaries of the rating system are set as 0 and 5, namely if
$x_{ik}$ is smaller than 0 we reset it as 0 and if it is larger than
5 we reset it as 5. According to Eq. (1), in the statistical level,
$\sigma_i\sim \zeta_i^2$.

Initially we set $\forall_i\xi_i=1$ and then apply the iterative
algorithm described in Eqs. (2) and (3). After we obtain the
convergent $\vec{q}$ and $\vec{\xi}$, we use standard deviation to
quantify to what extent our method can uncover the the intrinsic
qualities of objects:
\begin{equation}
\delta=\sqrt{\frac{1}{M}\sum_{l=1}^{M}(q_{l}-Q_{l})^{2}}.
\end{equation}
Clearly, a smaller $\delta$ corresponds to better algorithmic
performance. Besides, we use a correlation measure called
\emph{Kendall' Tau} \cite{Kendall1938} to judge whether our
algorithm has successfully revealed the hidden rank of users'
reputations. For two lists, $Y$ and $Z$, with length $L$, $\tau$ is
given as:
\begin{equation}
\tau=\frac{2}{L(L-1)}\sum_{i<j}\theta _{ij}~,~\theta
_{ij}=\texttt{sgn}[(Y_{i}-Y_{j})(Z_{i}-Z_{j})],
\end{equation}
where $\texttt{sgn}(x)=1$ for $x>0$, $\texttt{sgn}(x)=-1$ for $x<0$
and $\texttt{sgn}(x)=0$ for $x=0$. The value of $\tau$ ranges from
+1 (exactly the same order of the two lists $Y$ and $Z$) to -1
(completely reverse order of the two lists), and $\tau\approx 0$ for
uncorrelated lists. Clearly, a larger $\tau$ corresponds to better
algorithmic performance.

\begin{figure}
\begin{center}
\includegraphics*[scale=0.8]{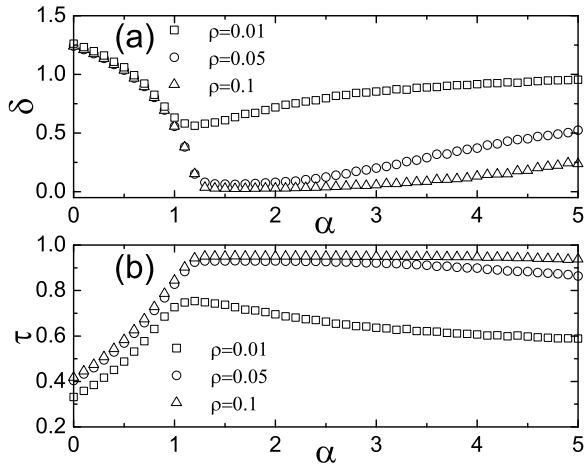}
\end{center}
\caption{$\delta$ and $\tau$ as functions of $\alpha$, where $Q$
obeys the power-law distribution $p(Q)\sim Q^{-1.5}$ and $\zeta$
obeys the uniform distribution. The squares, circles and triangles
represent the results for $\rho=0.01$, $\rho=0.05$ and $\rho=0.10$,
respectively. All data points are obtained by averaging 100
independent realizations. The results with $Q$ obeying the uniform
distribution are qualitatively the same, as thus are omitted here.}
\end{figure}

To our knowledge, there is no empirical analysis about the
quantitative distribution of people's judgements, so we simply
assume that $\zeta$ obeys a uniform distribution in the range
$[0,5]$. For the qualities of objects, we test on two kinds of
distributions: the uniform distribution and the power-law
distribution $p(Q)\sim Q^{-1.5}$. We adopt the latter distribution
because for many user-object bipartite systems the degrees of
objects are very heterogeneous \cite{Shang2009}, indicating that the
qualities of objects may be also heterogeneous. The value of $Q$ is
also restricted in the range $[0,5]$.

\begin{table}
\caption{Basic statistics of real data.}\label{tab1}
\begin{tabular}{ccccc}
\hline
Data Set & $M$ & $N$ & S & $\rho$ \\
\hline
Amazon &  10000 &  16311 & 189 & 0.0002 \\
\hline
Movielens & 3900 & 6040 & 74 & 0.0425 \\
\hline
\end{tabular}
\end{table}

Figure 1 and Figure 2 respectively report the algorithmic
performance with different distributions of object qualities.
Although the shapes of $\delta-\alpha$ curves and $\tau-\alpha$
curves in Fig. 1 and Fig. 2 are different in some details, both
figures clearly show the advantage of our algorithm. Compared with
the simple average (i.e., the case of $\alpha=0$), our algorithm can
provide considerably better evaluations on user reputation and
object quality. We next study the effects of rating density on
algorithmic performance. As shown in Fig. 3, the algorithm performs
better for denser data but the qualitative features do not change
for different $\rho$.

\section{Experimental Results}
In this section, we test our algorithm on two real data sets:
\emph{MovieLens} (http://www.grouplens.org/) and \emph{Amazon}
(http://www.amazon.com/). The former consists of 6040 users and 3900
movies, and the latter consists of 16311 users and 10000 books (the
Amazon data was collected from July 2005 to September 2005). All the
ratings on movies and books are discrete integers from 1 to 5. Since
in the real world, the users' reputations and objects' qualities
could never be exactly observed or quantified, we are not able to
test the algorithmic performance in a direct way. Instead, we first
select a subset of objects as benchmark ones that are known to be of
high quality, and then see whether our algorithm assigns in average
higher ranks to these benchmark objects than the simple average of
ratings. We apply the AUC statistics \cite{Hanely1982} to evaluate
our algorithm, which is the probability a randomly selected
benchmark object is assigned topper rank than a randomly selected
non-benchmark one, as
\begin{equation}
AUC=\frac{1}{S}\sum_i\frac{M-R_{i}}{M-S},
\end{equation}
where $S$ denotes the number of benchmark objects, $i$ runs over all
benchmark objects and $1\leq R_i \leq M$ is the rank of object $i$.
A completely random order of objects corresponds to $AUC=0.5$,
therefore, the degree to which $AUC$ exceeds 0.5 indicates how much
better the algorithm performs than pure chance. 74 movies winning
the \emph{Best Picture of Oscar Awards} and 189 books winning the
\emph{National Book Awards} are selected to be the benchmark objects
for MovieLens and Amazon, respectively. The basic statistics of real
data sets are shown in Table 1.

\begin{figure}
\begin{center}
\includegraphics*[scale=0.8]{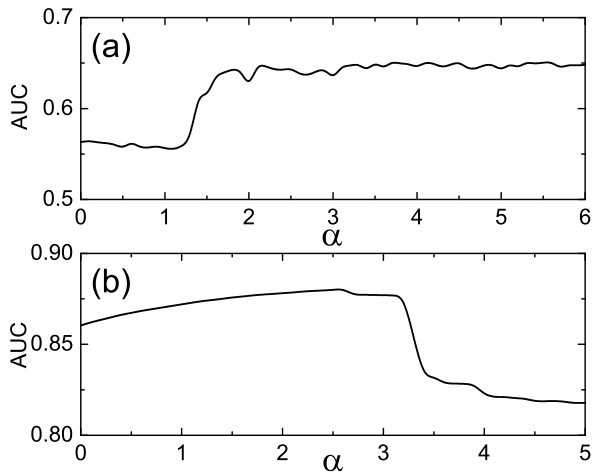}
\end{center}
\caption{ $AUC$ value as a function of $\alpha$ for Amazon (a) and
MovieLens (b). Results are obtained by averaging over 100
independent realizations since the objects with the same $q$ value
may be assigned different orders in different realizations. }
\end{figure}

Figure 4 reports the experimental results. Although the shapes of
$AUC-\alpha$ curves are different for MovieLens and Amazon (they are
also different from the artificial systems), our algorithm
outperforms the simple average in both two data sets. In accordance
with the results of artificial data, the sparser the ratings are,
the smaller the AUC is.

\section{Conclusion and Discussion}
As stated by Masum and Zhang \cite{Masum2004}, how to quantify
people's reputation is an urgent challenge in the Internet era. For
example, spammers intentionally produce noisy and evil information
that misleads our judgement, and the well-designed reputation
systems can dig out these nasty users or reduce their impacts. In
this paper, we focus on the bipartite rating systems, and design an
iterative refinement method to evaluate the users' reputations and
objects' qualities. According to both the artificially generated
data and the real data, our algorithm could considerably improve the
evaluation accuracy. In addition, the method adopted to test the
algorithm for real data (a similar method is reported very recently
in Ref. \cite{Radicchi2009}) suggests a good platform for the
quantitative competition of different ranking algorithms. To our
knowledge, although some reputation-based ranking algorithms have
been proposed previously
\cite{Yu2006,Laureti2006,Fouss2010,Kerchove2007}, no empirical
comparison between them has been reported yet, and it is not easy to
say one algorithm could beat another without a reasonable metric on
algorithmic performance for real data. Thanks to the increasing
number of available data sets and the metric suggested in this
paper, extensive comparison between various algorithms become
feasible \cite{Medo2010}, from which we hope the effectiveness and
efficiency of related algorithms can be largely improved in the near
future.

\begin{acknowledgements}
We acknowledge the GroupLens Research Group for MovieLens data and
Franti\v{s}ek Slanina for collecting the Amazon data. This work was
partially supported by the Future and Emerging Technologies
programme FP7-COSI-ICT of the European Commission through project
QLectives (Grant No. 231200) and the Swiss National Science
Foundation (Grant No. 200020-121848). TZ acknowledges the National
Natural Science Foundation of China under Grant Nos. 10635040,
60744003 and 60973069.
\end{acknowledgements}


\begin{thebibliography}{ref1}
\bibitem{Freeman2004} L. C. Freeman, \emph{The Development of Social Network Analysis: A Study in the
Sociology of Science} (Empirical Press, Vancouver, Canada, 2004).
\bibitem{Brin1998} S. Brin and L. Page, Comput. Netw. ISDN Syst.
{\bf 30}, 107 (1998).
\bibitem{Kleinberg1999} J. Kleinberg, J. ACM {\bf 46}, 604 (1999).
\bibitem{Walker2007} D. Walker, H. Xie, K.-K. Yan, and S. Maslov, J.
Stat. Mech. P06010 (2007).
\bibitem{Ding2009} Y. Ding, E. Yan, A. Frazho, and J. Caverlee, J.
Am. Soc. Inf. Sci. Technol. {\bf 60}, 2229 (2009).
\bibitem{Radicchi2009} F. Radicchi, S. Fortunato, B. Markines, and A.
Vespignani, Phys. Rev. E {\bf 80}, 056103 (2009).
\bibitem{Masum2004} H. Masum and Y.-C. Zhang, \emph{Manifesto for the
reputation society}, First Monday (5, July, 2004).
\bibitem{Massa2004} P. Massa and B. Bhattacharjee, Lect. Notes
Comput. Sci. {\bf 2995}, 221 (2004).
\bibitem{Ziegler2004} C.-N. Ziegler and G. Lausen, Lect. Notes
Comput. Sci. {\bf 3291}, 840 (2004).
\bibitem{Gupta2003} M. Gupta, P. Judge, and M. Ammar, \emph{A reputation system for peer-to-peer
networks}, in \emph{Proceedings of the 13th international workshop
on Network and operating systems support for digital audio and
video} (ACM Press, 2003, p. 144-152).
\bibitem{Wei2006} D. Wei, S.-B. Yang, and L.-T. Gao, \emph{Object Reputation Based Anti-Pollution P2P File Sharing System}, in \emph{Proceedings of the 1st IEEE International Conference on Digital Information Management} (IEEE
Press, 2006, p. 538-543).
\bibitem{Walsh2006} K. Walsh and E. G. Sirer, \emph{Experience with an Object Reputation System for Peer-to-Peer Filesharing}, in \emph{Proceedings of the 3rd Symposium on Networked Systems Design and Implementation}, 2006.
\bibitem{Resnick2002} P. Resnick and R. Zeckhauser, Adv. Appl.
Microeconomics {\bf 11}, 127 (2002).
\bibitem{Josang2007} A. J{\o}sang, R. Ismail, and C. Boyd, Decision
Support Syst. {\bf 43}, 618 (2007).
\bibitem{Buchegger2003} S. Buchegger and J.-Y. Le Boudec, EPFL IC
Technical Report IC/2003/50.
\bibitem{Yu2006} Y.-K. Yu, Y.-C. Zhang, P. Laureti, and L. Moret, Physica A, {\bf 371},
732 (2006).
\bibitem{Laureti2006} P. Laureti, L. Moret, Y.-C. Zhang, and Y.-K.
Yu, Europhys. Lett. {\bf 75}, 1006 (2006).
\bibitem{Fouss2010} F. Fouss, A. Achbany, and M. Saerens, \emph{A Probabilistic Reputation Model} (unpublished).
\bibitem{Laureti2005} P. Laureti, L. Moret, and Y.-C. Zhang, Physica
A {\bf 345}, 705 (2005).
\bibitem{Kerchove2007} C. de Kerchove and P. van Dooren, arXiv:
0711.3964.
\bibitem{Kendall1938} M. Kendall, Biometrika {\bf 30}, 81 (1938).
\bibitem{Shang2009} M.-S. Shang, L. L\"u, Y.-C. Zhang, and T. Zhou,
arXiv: 0909.4938.
\bibitem{Hanely1982} J. A. Hanely and B. J. McNeil, Radiology {\bf
143}, 29 (1982).
\bibitem{Medo2010} M. Medo, J. R. Wakeling, T. Zhou, L.-L. Jiang,
C.-H. Jin, and Y.-C. Zhang, \emph{Comparative study of
reputation-based ranking methods} (unpublished).

\end{thebibliography}
\end{document}